# On some constraints that limit the design of an invisibility cloak

Huanyang Chen[1,2], Xunya Jiang[1,3] and C. T. Chan[1]

[1]Department of Physics, Hong Kong University of Science and Technology,
Clear Water Bay, Kowlong, Hong Kong, China

[2]Institute of Theoretical Physics, Shanghai Jiao Tong University, Shanghai 200240, China

[3]Institute of Microsystem and Information Technology, CAS, Shanghai 200050, China

**Abstract:** Using the idea of transformation medium, a cloak can be designed to make a domain invisible for one target frequency. In this article, we examine the possibility to extend the bandwidth of such a cloak. We obtained a constraint of the band width, which is summarized as a simple inequality that states that limits the bandwidth of operation. The constraint originates from causality requirements. We suggest a simple strategy that can get around the constraint.

There has always been a keen interest in designing a coating that can reduce the scattering cross section of an object both for academic and practical reasons, and the idea of using "transformation media" to achieve invisibility has drawn great interest recently [1, 2, 3]. Similar ideas can be used to manipulate EM waves, such as field concentration [4] and field rotation [5]. It was noted that the invisibility cloak operates at one single frequency [2, 3] due to the causality constraints [6]. For example, the electromagnetic cloak at microwave frequencies works at 8.5GHz [2], while the optical cloaking is designed to function at $\lambda = 632.8 nm$ [3]. In this article, we see how we may obtain a reduced scattering cross section (invisibility is the ultimate limit) for a range of frequencies based on the idea of coordinate transformation [1]. We limit ourselves to 2D systems. In the following, we will use the term "dispersive cloak" to mean a cloak that works for a finite bandwidth with the frequency dispersion of the metamaterial taken into consideration.

The original 2D ideal cloak requires anisotropic permittivity and permeability of the form [7]:

$$\mu_r = \frac{r-a}{r}, \mu_\theta = \frac{r}{r-a}, \varepsilon_z = (\frac{b}{b-a})^2 \frac{r-a}{r} \quad (1)$$

This cloak is generated by compressing a cylindrical region r<b into a concentric cylindrical shell a<r<b by the mapping $r' = a + r\frac{b-a}{b}$. The cloaking is equivalent to the free space.

Let us consider a generalized problem. Suppose we have a perfect electrical conductor (PEC) cylinder with radius $r = r_0$ in the free space. If the wavelength of the incident wave is much larger than $r_0$, the scattering will be weak. We compress the concentric cylindrical shell region $r_0 < r < b$ into another shell $a < r < b$ with the coordinate transformation

$r = b - \frac{b-r_0}{f(b)-f(a)}(f(b) - f(r'))$. Here, $f(r)$ is a smooth function. The anisotropic

permittivity and permeability in the cloaking shell $a < r < b$ that can give an equivalent distortion of space are given by,

$$\mu_r = \frac{f(r) - f(a')}{rf'(r)}, \mu_\theta = \frac{rf'(r)}{f(r) - f(a')}, \varepsilon_z = (\frac{b - r_0}{f(b) - f(a)})^2 \frac{f'(r)(f(r) - f(a'))}{r} \quad (2)$$

where $f(a') = \frac{bf(a) - r_0 f(b)}{b - r_0}$. For the special case of $r_0 = 0$, $f(r) = r$ and $f'(r) = 1$, Eq. 2 reduces to Eq. 1. Notice that this mapping preserves the PEC boundary condition from $r = r_0$ to $r = a$.

If we choose in particular $f(r) = r$, then $a' = b\frac{a - r_0}{b - r_0}$. Eq. 2 becomes:

$$\mu_r = \frac{r - a'}{r}, \mu_\theta = \frac{r}{r - a'}, \varepsilon_z = (\frac{b - r_0}{b - a})^2 \frac{r - a'}{r} \quad (3)$$

Fig. 1 shows the equivalency of the PEC cylinder and the corresponding transformation media. The incident TE plane wave is from left to right in x-direction, whose frequency is 8.5GHz. Fig. 1a shows the scattering pattern of the transformation media in Eq. 3 with its inner radius $a$ = 27.1mm and the outer radius b = 58.9mm which are the same scale of the experiment cloak [2]. Note that PEC is at r=a. Fig. 1b shows the scattering pattern of the PEC cylinder with $r_0 = a/2$.

From Fig. 1a and 1b, the same scattering pattern can be observed at the region $r > b$, which demonstrates the equivalency mentioned above. The results are calculated using the commercial finite-element solver COMSOL MULTIPHYSICS.

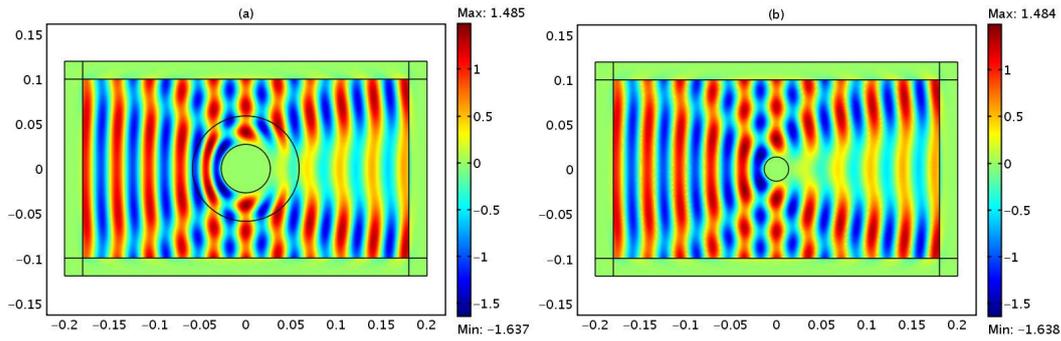

Fig. 1 (Color online) (a) Electric-field distribution in the vicinity of the transformation media. (b) Electric-field distribution in the vicinity of the PEC cylinder.

The principal refractive indices for Eq. 3 are [6],

$$n_r = \sqrt{\varepsilon_z \mu_\theta} = \frac{b - r_0}{b - a} \quad (4)$$

$$n_\theta = \sqrt{\varepsilon_z \mu_r} = \frac{b-r_0}{b-a}\frac{r-a'}{r} = \frac{r(b-r_0)-b(a-r_0)}{r(b-a)} \quad (5)$$

We can keep these two principal refractive indices unchanged by several methods of "reduction" [2, 3], which lumps all spatial dependence to just $\mu_r$. For instance, we can have,

$$\mu_r = (\frac{r-a'}{r})^2, \mu_\theta = 1, \varepsilon_z = (\frac{b-r_0}{b-a})^2 \quad (6)$$

We note that $a'$ is a function of $r_0$. Such "reduction" makes the anisotropic metamaterial much easier to realize in practice. We will see later that such a reduction is actually necessary for expanding the bandwidth of operation.

To achieve an invisible cloak, we would like to set $r_0 = 0$. We can do it for one frequency, but from Eq. (4)-(6), we see that we if we choose to set $r_0 = 0$ for one frequency, say $\omega_a$, we cannot set $r_0 = 0$ for another frequency $\omega_a + \Delta\omega$, since that would require that we have frequency-independent permittivity and permeability. For the reason that the metamaterials are intrinsically dispersive, the parameter $r_0$ is always a function of frequency, so that $r_0 = r_0(\omega)$.

Our task is to construct a system such that $r_0(\omega)$ is as small as possible for a certain frequency range. Suppose that, in a dispersive shell (the dispersion relation is known), we can find such a frequency range [$\omega_a$ - $\omega_b$], such at $\omega_a$ the cloaking with { $\varepsilon_z(\omega_a), \mu_r(\omega_a)$ and $\mu_\theta(\omega_a)$ } satisfies the best-possible case which corresponds to $r_0 = 0$ (same as the ideal case or its reduction[7]), but at $\omega_b$ the cloaking with { $\varepsilon_z(\omega_a), \mu_r(\omega_a)$ and $\mu_\theta(\omega_a)$ } satisfies the Eq. 3 or Eq. 6 with $r_0 \neq 0$ but still a small value. In this frequency range, ($\omega_b$ is close to $\omega_a$), we can assume there is a continuous function that relates every $\omega$ value and every $r_0$ value. In other words, we can define $r_0$ as the function of $\omega$, say $r_0 = r_0(\varepsilon_z(\omega), \mu_r(\omega), \mu_\theta(\omega)) \equiv r_0(\omega)$, if the dispersion property of the material is given. With this, the relation between the frequency range (also the dispersion of material) with the geometry of transformation can be constructed.

Then we need to find out the dispersion relation in the cloak shell ($a < r < b$) as it is an anisotropic medium which is different from the isotropic one. We need to find out the index ellipsoid of the metamaterial, and make sure that the group velocity is less than $c$ in every direction. We seek plane wave solutions with the real wave vector $k = \frac{\omega}{c}\kappa$ (or $\kappa = \frac{c}{\omega}k$). As the $\varepsilon$ and $\mu$ are the same symmetric tensors for Eq. 3, which we denote them as **n**, we can obtain the index ellipsoid by the equation [8]:

$$(\kappa n \kappa - \det(n)) = 0 \quad (7)$$

For a more explicit relation, we have [7, 8]:

$$\begin{aligned}
(\kappa n \kappa - \det(n)) &= \kappa_x^2 n_{11} + 2\kappa_x \kappa_y n_{12} + \kappa_y^2 n_{22} - n_{33} \\
&= \kappa_x^2 (\mu_r \cos^2\theta + \mu_\theta \sin^2\theta) + 2\kappa_x \kappa_y (\mu_r - \mu_\theta)\sin\theta\cos\theta \\
&\quad + \kappa_y^2 (\mu_r \sin^2\theta + \mu_\theta \cos^2\theta) - \varepsilon_z \\
&= \mu_r (\kappa_x^2 \cos^2\theta + 2\kappa_x \kappa_y \sin\theta\cos\theta + \kappa_y^2 \sin^2\theta) \\
&\quad + \mu_\theta (\kappa_x^2 \sin^2\theta - 2\kappa_x \kappa_y \sin\theta\cos\theta + \kappa_y^2 \cos^2\theta) - \varepsilon_z \\
&= \mu_r (\kappa_x \cos\theta + \kappa_y \sin\theta)^2 + \mu_\theta (\kappa_x \sin\theta - \kappa_y \cos\theta)^2 - \varepsilon_z = 0
\end{aligned} \quad (8)$$

Since for the $\mu_r \mu_\theta = 1$, and from Eq. 4 and Eq. 5, we can write the above equation as,

$$\frac{(\kappa_x \cos\theta + \kappa_y \sin\theta)^2}{\varepsilon_z \mu_\theta} + \frac{(\kappa_x \sin\theta - \kappa_y \cos\theta)^2}{\varepsilon_z \mu_r} - 1 = \frac{\kappa_r^2}{n_r^2} + \frac{\kappa_\theta^2}{n_\theta^2} - 1 = 0 \quad (9)$$

where 
$$\begin{aligned}
\kappa_r &= \kappa_x \cos\theta + \kappa_y \sin\theta \\
\kappa_\theta &= \kappa_x \sin\theta - \kappa_y \cos\theta \\
k_r &= k_x \cos\theta + k_y \sin\theta \\
k_\theta &= k_x \sin\theta - k_y \cos\theta
\end{aligned}$$
. This dispersion relation is also valid for the reduction of cloaks (Eq. 6).

Define the function $f(k_x, k_y, \omega)$ as:

$$f(k_x, k_y, \omega) = \frac{\frac{c^2}{\omega^2}(k_x \cos\theta + k_y \sin\theta)^2}{\varepsilon_z \mu_\theta} + \frac{\frac{c^2}{\omega^2}(k_x \sin\theta - k_y \cos\theta)^2}{\varepsilon_z \mu_r} - 1$$

$$= \frac{\frac{c^2}{\omega^2}k_r^2}{\varepsilon_z \mu_\theta} + \frac{\frac{c^2}{\omega^2}k_\theta^2}{\varepsilon_z \mu_r} - 1 = \frac{\kappa_r^2}{n_r^2} + \frac{\kappa_\theta^2}{n_\theta^2} - 1 = 0 \quad (10)$$

then,

$$\begin{aligned}
\partial f / \partial k_x &= \frac{2\frac{c^2}{\omega^2}(k_x \cos\theta + k_y \sin\theta)\cos\theta}{n_r^2} + \frac{2\frac{c^2}{\omega^2}(k_x \sin\theta - k_y \cos\theta)\sin\theta}{n_\theta^2} \\
&= \frac{2\frac{c}{\omega}\kappa_r \cos\theta}{n_r^2} + \frac{2\frac{c}{\omega}\kappa_\theta \sin\theta}{n_\theta^2}
\end{aligned} \quad (11)$$

$$\partial f / \partial k_y = \frac{2\frac{c^2}{\omega^2}(k_x \cos\theta + k_y \sin\theta)\sin\theta}{n_r^2} - \frac{2\frac{c^2}{\omega^2}(k_x \sin\theta - k_y \cos\theta)\cos\theta}{n_\theta^2} \quad (12)$$

$$= \frac{2\frac{c}{\omega}\kappa_r \sin\theta}{n_r^2} - \frac{2\frac{c}{\omega}\kappa_\theta \cos\theta}{n_\theta^2}$$

$$-\partial f / \partial \omega = 2\frac{c^2}{\omega^3}(\frac{k_r^2}{n_r^2} + \frac{k_\theta^2}{n_\theta^2}) + 2\frac{c^2}{\omega^2}(\frac{k_r^2}{n_r^3}\frac{dn_r}{d\omega} + \frac{k_\theta^2}{n_\theta^3}\frac{dn_r}{d\omega})$$

$$= 2\frac{c^2}{\omega^3}\frac{k_r^2}{n_r^3}(n_r + \omega\frac{dn_r}{d\omega}) + 2\frac{c^2}{\omega^3}\frac{k_\theta^2}{n_\theta^3}(n_\theta + \omega\frac{dn_\theta}{d\omega}) \quad (13)$$

$$= 2\frac{c^2}{\omega^3}\frac{k_r^2}{n_r^2}\frac{m_r}{n_r} + 2\frac{c^2}{\omega^3}\frac{k_\theta^2}{n_\theta^2}\frac{m_\theta}{n_\theta} = 2\frac{1}{\omega}\frac{\kappa_r^2}{n_r^2}\frac{m_r}{n_r} + 2\frac{1}{\omega}\frac{\kappa_\theta^2}{n_\theta^2}\frac{m_\theta}{n_\theta}$$

From the definition of the group velocity [9], we have,

$$\vec{v}_g = v_{gx}\hat{i} + v_{gy}\hat{j} = (\partial f/\partial k_x \hat{i} + \partial f/\partial k_y \hat{j})/(-\partial f/\partial \omega) \quad (14)$$

The causality needs the following condition:

$$v_g^2 = ((\partial f/\partial k_x)^2 + (\partial f/\partial k_y)^2)/(-\partial f/\partial \omega)^2 \leq c^2 \quad (15)$$

That is:

$$\frac{(\frac{\kappa_r^2}{n_r^4} + \frac{\kappa_\theta^2}{n_\theta^4})}{(\frac{\kappa_r^2}{n_r^2}\frac{n_r + \omega\frac{dn_r}{d\omega}}{n_r} + \frac{\kappa_\theta^2}{n_\theta^2}\frac{n_\theta + \omega\frac{dn_\theta}{d\omega}}{n_\theta})^2} < 1 \quad (16)$$

with $\frac{\kappa_r^2}{n_r^2} + \frac{\kappa_\theta^2}{n_\theta^2} = 1$. We can write it in the form of real wave vector:

$$(\frac{k_r^2}{n_r^4} + \frac{k_\theta^2}{n_\theta^4})\frac{\omega^2}{c^2} < (\frac{k_r^2}{n_r^2}\frac{n_r + \omega\frac{dn_r}{d\omega}}{n_r} + \frac{k_\theta^2}{n_\theta^2}\frac{n_\theta + \omega\frac{dn_\theta}{d\omega}}{n_\theta})^2 \quad (17)$$

with $\frac{k_r^2}{n_r^2} + \frac{k_\theta^2}{n_\theta^2} = \frac{\omega^2}{c^2}$.

We introduce an auxiliary angle such as $k_r = \frac{\omega}{c}n_r \sin\tau$ and $k_\theta = \frac{\omega}{c}n_\theta \cos\tau$ to denote the components of the real wave vector in different directions.

There are two necessary conditions in order that Eq. 17 can be satisfied,

$$v_g = \frac{c}{n_\theta + \omega \frac{dn_\theta}{d\omega}} < c \quad \text{when} \quad k_\theta = \frac{\omega}{c} n_\theta \ (k_r = 0) \quad \text{and} \quad v_g = \frac{c}{n_r + \omega \frac{dn_r}{d\omega}} < c \quad \text{when}$$

$$k_r = \frac{\omega}{c} n_r \ (k_\theta = 0).$$

That is:

$$n_r + \omega \frac{dn_r}{d\omega} = \frac{b - r_0}{b - a} - \frac{1}{b - a} \omega \frac{dr_0}{d\omega} > 1 \quad (18)$$

$$n_\theta + \omega \frac{dn_\theta}{d\omega} = \frac{r(b - r_0) - b(a - r_0)}{r(b - a)} + \frac{b - r}{r(b - a)} \omega \frac{dr_0}{d\omega} > 1 \quad (19)$$

The above two inequalities will result in two conditions conflicts with each other for all frequencies: $\omega \frac{dr_0}{d\omega} < a - r_0$, $\omega \frac{dr_0}{d\omega} > a - r_0$.

The result shows that no analytic function $r_0(\omega)$ that can satisfy the causality requirement. In order words, the transformation media equations given by Eq 3 can only be realized in one single frequency, and cannot be extended to a finite bandwidth (no matter how small is the bandwidth) even if we allow $r_0$ to be a function of frequency. Changing the material to the "reduced" form will not help, since Eq. 6 has the same refractive index ellipsoid as Eq. 3. In order to construct a cloak that works for a range of frequencies, we need to make further modification. One possibility is to take

$$\mu_r = \frac{r - a'}{r}, \mu_\theta = \frac{r}{r - a'}, \varepsilon_z = (\frac{b - \bar{r}}{b - a})^2 \frac{r - a'}{r} \quad (20)$$

with its corresponding reduction:

$$\mu_r = (\frac{r - a'}{r})^2, \mu_\theta = 1, \varepsilon_z = (\frac{b - \bar{r}}{b - a})^2 \quad (21)$$

where $\bar{r}$ is a constant.

The principal refract indexes are then as

$$n_r = \sqrt{\varepsilon_z \mu_\theta} = \frac{b - \bar{r}}{b - a} \quad (22)$$

$$n_\theta = \sqrt{\varepsilon_z \mu_r} = \frac{b - \bar{r}}{b - a} \frac{r - a'}{r} \quad (23)$$

$n_r$ is not function of frequency but $n_\theta$ still is. From Eq. 17, we can obtain a constraint:

$$(\frac{n_\theta^2}{n_r^2} \sin^2 \tau + \cos^2 \tau) \leq (n_\theta \sin^2 \tau + (n_\theta + \omega \frac{dn_\theta}{d\omega}) \cos^2 \tau)^2 \quad (24)$$

Then we define two functions of $\tau$:

$$f_1(\tau) = \sqrt{\frac{n_\theta^2}{n_r^2} \sin^2 \tau + \cos^2 \tau}$$

$$f_2(\tau) = n_\theta \sin^2\tau + (n_\theta + \omega\frac{dn_\theta}{d\omega})\cos^2\tau \quad (25)$$

For any direction that is parameterized by a particular value of $\tau$, $f_2(\tau)$ should be always larger than $f_1(\tau)$. This condition is not always satisfied. For example, if $n_\theta = 0$, then, $f_1(\tau) = \cos\tau$ and $f_2(\tau) = (\omega\frac{dn_\theta}{d\omega})\cos^2\tau$. With $(\omega\frac{dn_\theta}{d\omega}) = 1.5$ for example we plot $f_1(\tau)$ and $f_2(\tau)$ in Fig. 2. We can see that for some directions the causality is broken. So one necessary requirement is that $n_\theta \neq 0$. This is an important constraint that is not satisfied by the original ideal cloak or its reduction [7], $n_\theta(r=a) = 0$.

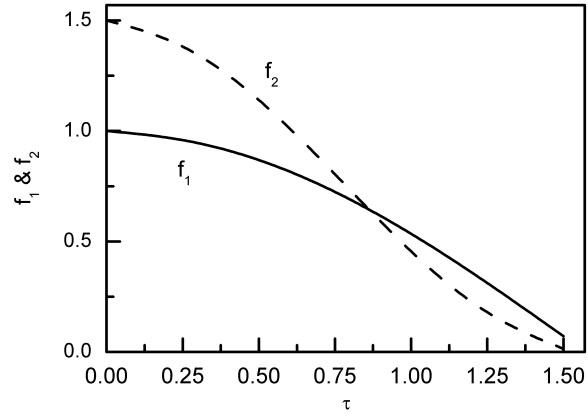

Fig. 2 The dependence of $f_1(\tau)$ and $f_2(\tau)$ with $\tau$. When $0 < \tau < 0.846$, $f_1(\tau) < f_2(\tau)$; when $0.846 < \tau < \pi/2$, $f_1(\tau) > f_2(\tau)$.

A potential dispersive cloak from Eq. 21 could be designed base on the causality constraint Eq. 24. However, we can extract some important intrinsic physical information for the direction $k_r = 0$:

$$n_\theta + \omega\frac{dn_\theta}{d\omega} = \frac{b-\bar{r}}{b-a}(\frac{r(b-r_0)-b(a-r_0)}{r(b-r_0)}) + \frac{b(b-a)}{r(b-r_0)^2}\omega\frac{dr_0}{d\omega}) > 1 \text{ or}$$

$$\omega\frac{dr_0}{d\omega} > (\frac{b-a}{b-\bar{r}} - \frac{r(b-r_0)-b(a-r_0)}{r(b-r_0)})\frac{r(b-r_0)^2}{b(b-a)} \quad (26)$$

Since the maximum of $(\frac{b-a}{b-\bar{r}} - \frac{r(b-r_0)-b(a-r_0)}{r(b-r_0)})\frac{r(b-r_0)^2}{b(b-a)}$ is $\frac{b}{b-\bar{r}}a$, so that Eq. 26 can be rewritten as:

$$\omega \frac{dr_0}{d\omega} > \frac{b}{b-\bar{r}} a \quad (27)$$

Since $\bar{r}$ is much smaller than $b$, at last we have a clean and meaningful inequality:

$$\omega \frac{dr_0}{d\omega} > a \quad \text{or} \quad \frac{\Delta \omega}{\omega} < \frac{\Delta r_0}{a} \quad (28)$$

This condition is a necessary condition for a dispersive cloak. This "invisibility condition" sets a limit on the band width of the type of "reduced" transformation media as specified by Eq. 21. The physical meaning of this condition is very clear. If we hope to have to be "invisible" in a larger frequency range for the cloaking based on the geometry transformation, we must have more tolerance of $r_0$ (the cross-section is larger, and easier to be observed).

In conclusion, we have shown that a generalized form of the transformation media can be realized at one single frequency, and cannot be made to be compatible with causality over an extended range of frequency, no matter how small that range is. However, by a simple adaptation, the "reduced" transformation media can be made to be compatible with causality, and we found that the reduced media obeys a simple "the invisibility condition", which limits the band width.

This work was supported by Hong Kong Central Allocation grant HKUST3/06C. Computation resources are supported by Shun Hing Education and Charity Fund.